\documentclass{JHEP3}%

\usepackage{subfigure}%
\usepackage{graphicx}%
\usepackage{enumerate}%
\usepackage{amsmath}%

\title{The dangers of extremes}

\author{Donald Marolf  \\
Physics Department, UCSB, Santa Barbara,  \\
CA 93106, USA \\\texttt{marolf@physics.ucsb.edu} \\
 \\
{\rm Essay written for the Gravity Research Foundation 2010 Awards for Essays on Gravitation} \\
\\
Submission date: March 28, 2010}

\abstract{
While extreme black hole spacetimes with
smooth horizons are known at the level of mathematics, we argue that the horizons of physical extreme black holes are effectively singular.  Test particles encounter a singularity the
moment they cross the horizon, and only objects with significant back-reaction can fall across a smooth (now non-extreme) horizon.  As a result, classical interior solutions for extreme
black holes are theoretical fictions that need not be reproduced by any quantum mechanical model. This observation suggests that significant quantum effects might be visible outside extreme or nearly extreme black holes. It also suggests that the microphysics of such black holes may be very different from that of their Schwarzschild cousins.}

\date{Submission date: March 2009}

\begin{document}

\FIGURE {\label{ExRNfig}
\includegraphics[width=1.5in]{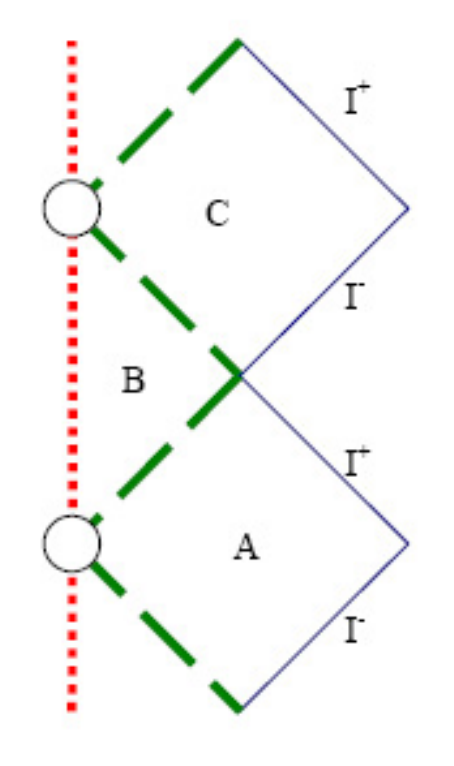} \caption{The conformal
diagram for an extreme RN  black hole.  Future and
past null infinity ($I^\pm$) are shown for two asymptotic regions, A
and C.  The dotted line is a time-like singularity and the dashed
lines are Cauchy horizons.  The two circles are internal infinities
lying at the ends of infinite throats.}}

\section{Introduction and review}
\label{intro}

For reasons related to the strict vanishing of their temperature, extreme black holes are generally considered impossible to form in classical physics.  While any non-extremality parameter can be made small with enough care, it cannot be brought to zero in finite time.  Yet precisely extreme black holes can in principle be formed through quantum processes.  For example, extreme Reissner-Nordstr\"om (RN) black holes can be pair created from the vacuum by strong electric and magnetic fields \cite{FGGT}.  Extreme magnetically-charged such black holes may even be stable in our universe.  More generally, in theories of gravity with appropriate supersymmetries, Hawking radiation can cause charged black holes to decay to extremality in finite time.  In this sense precisely extreme black holes can be said to be ``physical,'' at least in certain theories\footnote{Since we know of no mechanism to stabilize extreme Kerr against quantum mechanical radiation of gravitons, this essay will focus on spherically-symmetric extreme black holes.  We also take ``black holes'' to have compactly-generated horizons.}.

In the supersymmetric context, extreme solutions have provided a marvelous
laboratory in which to explore fundamental issues.  In particular, within string theory a detailed accounting of the Bekenstein-Hawking entropy has been given in terms of D-brane
configurations for many extreme and nearly extreme black holes following \cite{SV}.  We also mention the research program associated with Mathur \cite{Mathur} which attempts to replace black holes by collections of horizon-free geometries, and which has also focused on extremal settings.

Such works emphasize extreme black holes for which the maximal analytic extension of the stationary exterior has a smooth horizon
where curvatures are small.  Typical examples resemble extreme Reissner-Nordstr\"om,
whose conformal diagram is shown in figure
\ref{ExRNfig}.  It may thus appear natural to ask whether
these microscopic models also describe the region marked B inside
the horizon.  This can be a
large region in which curvatures are small so
long as one stays well away from the singularity.

\FIGURE{
\includegraphics[width=1.5in]{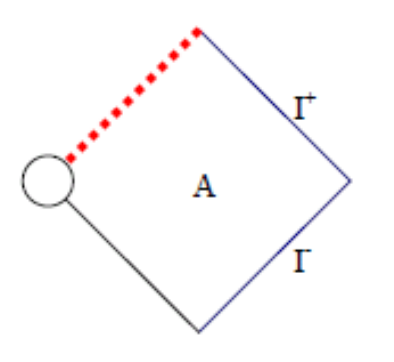} \caption{The effective conformal
diagram for physical extreme RN black holes.  The dotted line is a null
singularity.}  \label{Exinstab}}

Nevertheless, we argue below that physical extreme black holes simply have no analogue of region B.  Instead, although curvatures are small everywhere outside the horizon, test particles encounter a curvature singularity the instant they enter the black hole.  The physical extreme spacetime is better described by figure \ref{Exinstab}, in which the horizon has been replaced by a null singularity.  Infalling objects that produce large back-reaction effects may experience a smooth horizon, but one which is now non-extreme and the region they enter has properties quite different from region B.

Our argument relies on the mass-inflation instability of the inner horizon for non-extreme RN black holes, which we now review.
The conformal diagram
for the analytic extension of static non-extreme RN in shown in figure
\ref{NExRNfig}.  Recall that region $A_3$ is separated from
regions $B_1,B_2$ by Cauchy horizons, which are also the inner horizons
of the black hole.  One sees from the
diagram that initial data in $A_1,A_2$ does indeed determine
evolution in region $A_3$, but not in $B_1,B_2$.

\FIGURE
{\label{NExRNfig}
\includegraphics[width=3.5in]{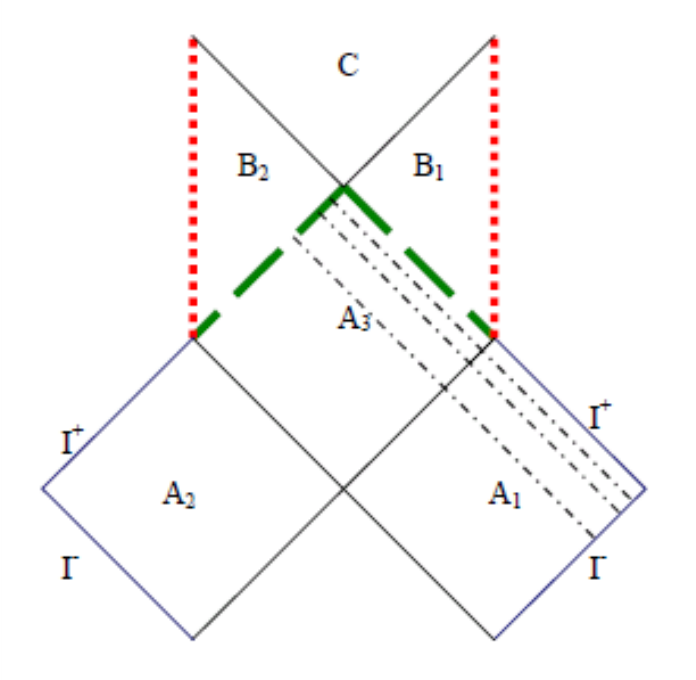}
\caption{ The analytic extension of a
non-extreme RN black hole. Again, the dotted
vertical lines are timelike singularities and the dashed lines are
Cauchy horizons.  The fine diagonal dash-dotted lines represent a series of
constant-phase surfaces for a radial ingoing spherical wave. The wave undergoes an infinite number of oscillations below $I^+$, and thus below the Cauchy horizon.
But the proper distance (or affine parameter) across the wavefronts
to the Cauchy horizon is finite.  Thus, any observer falling through
the Cauchy horizon would experience an infinite number of
oscillations in a finite time; i.e., the radiation has been
infinitely blueshifted
 } }

Recall also the infinite blueshift between $I^+$ and the Cauchy horizon, as illustrated in figure \ref{NExRNfig}. It was suggested long ago by Penrose \cite{Penrose} that this
blueshift could make the Cauchy horizon unstable.  Tiny bits of
(e.g.) gravitational radiation produced by an infalling object would
lead to a divergence, replacing the Cauchy horizon with a singularity.  This scenario was
confirmed by Poisson and Israel \cite{PI} and rigorously demonstrated
by Dafermos \cite{Dafermos}, though the presence of some right-moving perturbation also turns out to be important.

The full picture \cite{BradySmith} appears to be that shown in
figure \ref{instab}. Any perturbation triggers an instability which
forms a singularity.  One piece of the singularity is null
and simply replaces the would-be Cauchy horizon.   Along this null singularity the area
of the (singular) null congruence shrinks to zero, where it joins a
spacelike singularity.  Similar results appear to hold for
rotating black holes \cite{rot}, though the non-spherical setting is
more difficult to investigate.

\FIGURE
{ \label{instab}
\includegraphics[width=3in]{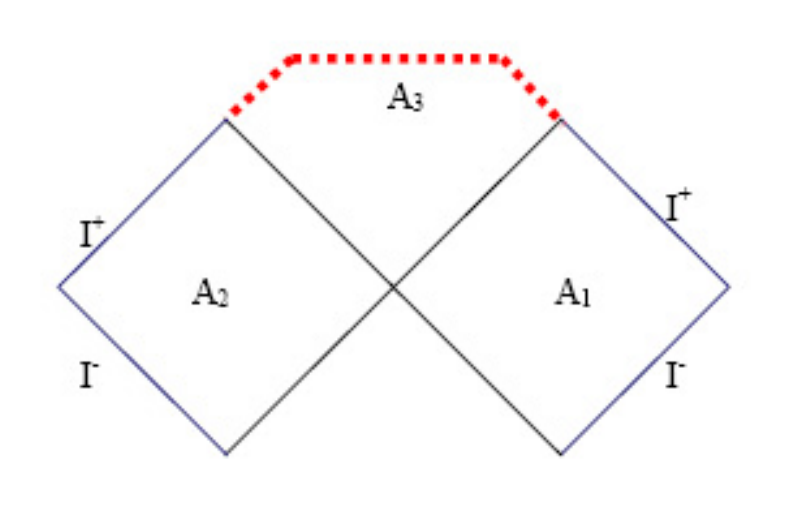}
\caption{A perturbed non-extreme RN black hole.
The perturbation triggers an instability which generates the
singularity indicated by the dotted line. The singularity may have both null and spacelike parts, though for a black hole with two asymptotic regions there need be no spacelike part if the perturbation is sufficiently weak.  No Cauchy horizons remain. } }

\section{Falling into extreme black holes}
\label{ext}

It is interesting to ask how far the new singularity of figure \ref{instab} lies inside the horizon, especially in the extreme limit where the inner horizon (at area-radius $r_-$) of the unperturbed hole coincides with the outer horizon (at area-radius $r_+$). For simplicity, we concentrate on black holes with two asymptotic regions as shown in figures \ref{NExRNfig} and \ref{instab}.  Black holes that form by collapse should behave similarly, at least as probed by observers who fall into the black hole long after it has formed.

To give meaning to the above question, consider a freely falling observer in the unperturbed spacetime of figure \ref{NExRNfig} who enters from region $A_1$ with energy per unit mass $\gamma = - U \cdot \partial_t$, where $U^a$ is the observer's 4-velocity and $\partial_t$ is the usual static Killing field of the exterior.  This observer requires a proper time $\Delta \tau$ to fall from $r_+$ to $r_-$ where
\begin{equation}
\label{dt}
\Delta \tau = \frac{r_+ - r_-}{\gamma} + {\cal O}\left((r_+ - r_-)^3/r_+^2 \gamma^3\right)
\end{equation}
with $c=1$.

Now consider a freely-falling observer satisfying the same initial conditions in the perturbed RN spacetime of figure \ref{instab}.  No matter how small the perturbation at $r_+$, before the proper time (\ref{dt}) elapses for our observer this perturbation must grow large, else the original cauchy horizon would remain.  But once the perturbation is large, the spacetime quickly evolves toward a singularity.  Our infalling observer must therefore reach the singularity in a time of order (\ref{dt}) after crossing the horizon.  Again, this conclusion holds {\it no matter how small the initial perturbation}, though in practice there there will of necessity be perturbations of some minimal size set by quantum fluctuations.

\FIGURE
{ \label{collapse}
\includegraphics[width=1.4in]{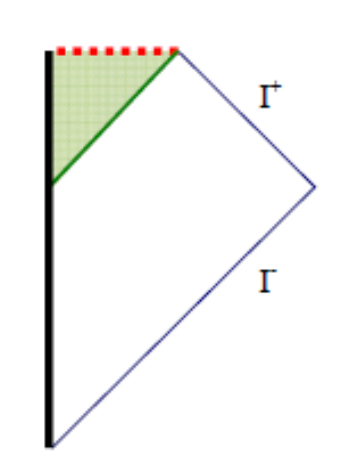}
\caption{The conjectured diagram for extreme black holes formed by decay of non-extreme black holes with a single asymptotic region.  The conformal structure resembles that of Schwarzschild.  However, the time-translation symmetry of the exterior will be strongly broken in the interior (shaded) so that the time required for $\gamma =1$ observers to fall from the horizon to the singularity will vanish in the late-time limit.  Observers who fall in long after formation will effectively experience fig. \ref{Exinstab}.}
}

For observers with fixed $\gamma$, the time (\ref{dt}) vanishes in the extreme limit $r_- \rightarrow r_+$.  Their experiences in this limit are thus described by figure \ref{Exinstab}, in which the singularity coincides with the outer horizon.  In this sense, figure \ref{Exinstab} describes the extreme limit of a physical RN black hole\footnote{Strictly speaking, the internal infinity of figure \ref{Exinstab} is removed by whatever process creates the extreme black hole; e.g., decay from non-extremality by Hawking radiation.  One might say that the infinite throat has been capped.  However, this cap recedes with time so that an internal infinity in some sense emerges after a long time has passed.}.

Suppose now that one is given a precisely extreme black hole, created perhaps through one of the quantum processes mentioned in the introduction.  Any observer falling into the black hole will carry with them a small amount of energy.  Because the horizon is compactly-generated, any such perturbation will make the black hole non-extreme.  It follows that the experience of any observer (with small but non-zero back-reaction) falling into an extreme black hole is given by the extreme limit of their experience falling into non-extreme black holes\footnote{There is no corresponding argument for non-compact extreme horizons such as the Poincar\'e horizon of anti-de Sitter space.  In such cases, non-extreme versions of the original horizon may not even satisfy the same boundary conditions and there is no reason to expect continuity.}.

The details of this experience are determined by the observer's back-reaction.  The larger the resulting non-extremality, the more proper time may pass before encountering a singularity.  However, defining the test particle limit as the limit of zero back reaction, one may say that test particles experience a singularity the moment they cross the original horizon.

As noted above, the situation may be somewhat different if the extreme black hole forms from decay of a single-asymptotic-region black hole, but only as probed by observers who fall in soon after the black hole has formed.  Late time observers falling into the black hole must again quickly encounter a singularity.  We expect this situation to be described by figure \ref{collapse}, which as described in the caption approximates figure \ref{Exinstab} from the viewpoint of late-time observers.  In particular, the classical interior solution described by region B of figure \ref{ExRNfig} is a theoretical fiction that does not describe the experience of any observer.

It is interesting to ponder the implications for quantum discussions of black holes.  First and foremost, the presence of strong curvatures (and strong curvature gradients) {\it immediately} inside the horizon suggests that significant quantum effects might possibly be visible outside extreme black holes, perhaps in scattering processes involving very low energy quanta as these events would be associated with minimal back-reaction.  While we have nothing specific to propose at this time, there may be a window of opportunity,  especially for non-perturbative effects.   It is also conceivable that such non-perturbative quantum effects could provide large corrections to the entropy of extreme black holes\footnote{Since the spacetime outside is not significantly disturbed, one would not expect large perturbative quantum effects.  Indeed, perturbative analyses of quantum corrections to extreme black holes indicate no such effects \cite{lowe}.}.

In addition, the above observation alters our expectations for microscopic descriptions of extreme black holes. This is good news for many proposed models.  It is not hard to imagine that an object's collision with either a stack of D-branes or with the Planck-curvature fuzz suggested by Mathur's program \cite{Mathur} would destroy the object in a manner that might resemble the collision with a singularity.    However, since the horizons of extreme black holes are so different from those of their Schwarzschild cousins, this victory is somewhat hollow.  The challenge for such models is merely postponed to the difficult regime of highly excited states, with large energies above extremality, where the models must somehow reproduce the gentle experience of an observer crossing a large smooth horizon.  Since the macrophysics is very different in this regime, appropriate microscopic models may be very different as well.

\section*{Acknowledgements}

It is a pleasure to thank Patrick Brady, Gary Horowitz, Harvey Reall, and Eric Poisson for numerous discussions over many years of both extreme and non-extreme black holes.
This work was supported in
part by the National Science Foundation under Grant No PHY08-55415,
and by funds from the University of California.

\end{document}